# Mathematics of Predicting Growth

Ron W Nielsen[1]

Environmental Futures Research Institute, Gold Coast Campus, Griffith University, Qld, 4222, Australia

October, 2015

Mathematical methods of analysis of data and of predicting growth are discussed. The starting point is the analysis of the growth rates, which can be expressed as a function of time or as a function of the size of the growing entity. Application of these methods is illustrated using the world economic growth but they can be applied to any other type of growth.

**Introduction**

The usual method of data analysis is based on the examination of changes in the *size*, *S*, of the growing entity. This simple method can be extended to the examination of related distributions such as $1/S$ or $\ln S$, which give a new insight into the interpretation of data.

The aim of any data analysis should be always to look for the simplest mathematical descriptions, which can then be used to explain the mechanism of growth. Different representations of data can allow not only for looking at them from a new perspective but also for finding a simple way of their analysis. For instance, the analysis of the historical economic growth or the growth or human population is difficult when using the size of the Gross Domestic Product (GDP) or the size of the population because both of them increase hyperbolically and such a growth is often interpreted incorrectly. However, their analysis is trivial (Nielsen, 2014) when using their reciprocal values, $1/S$, of data.

The aim of the discussion presented in this publication is to explain an additional way of analysis of data based on the examination of the *growth rates*. Data are of primary importance in scientific investigations and the more we know how to analyse them, the more successful we can be with their interpretation.

The analysis of data usually starts with their display. If the data vary over a large range of values, then displaying them by using linear scales of reference is not helpful because while the large values are shown clearly the small values are hard to interpret. In such a case it is a standard procedure to present them using semilogarithmic frames of reference, because we can then study simultaneously the features characterising the small and the large values of data. If in addition the range of the independent variable, such as time, is also large we could use the logarithmic scales for both axes of reference.

Semilogarithmic scales of reference can be also used to identify exponential growth because such a growth is then represented by a straight line. Likewise, displaying the reciprocal values of data, $1/S$, can be used to identify the first-order hyperbolic growth because it is then represented by a decreasing straight line. However, both the semilogarithmic display and the display of the reciprocal values of data can also help in studying fine details of growth.

---

[1] AKA Jan Nurzynski, r.nielsen@griffith.edu.au; ronwnielsen@gmail.com; http://home.iprimus.com.au/nielsens/ronnielsen.html

Such visual examinations of data can and even should be followed by their mathematical analysis with the aim of finding a mathematical description of data. The important point to remember is that such analysis should be *as simple as possible*, because our ultimate aim is to explain the mechanism of growth, and if we use complicated descriptions we can expect complicated explanations, which could be unconvincing. The fundamental principle of scientific investigations is to look for the simplest solutions and explanations. Unfortunately, this principle is often forgotten and it appears that there is a continuing desire to *construct* complicated formulae (e.g. Artzrouni & Komlos, 1985; Galor, 2005a, 2011; Johansen, & Sornette, 2001; Khaltourina & Korotayev, 2007; Korotayev, 2005; Korotayev, Malkov & Khaltourina, 2006a, 2006b; Lagerlöf, 2003) maybe to impress the reader or to have the paper published. Ironically, however, such complicated formulae are often not even tested by data.

We should understand the difference between *constructing* and *deriving* mathematical formulae. The process of constructing is guided entirely by creative imagination. Various elements are added here and there just because they look good or because they appear to do what we want them to do. It is usually a translation of concepts into a mathematical language, but incorrect concepts remain incorrect even if dressed up in mathematical gowns. Such exercises are particularly puzzling when it is already well known that a studied process can be described using a much simpler mathematical expression.

*Constructing* complicated mathematical formulae but failing to test them by accessible data is not helpful. Devoting 20 years of one's life (Baum, 2011) into developing a complicated theory (Galor, 2005a, 2011) but failing to see that its fundamental postulates are contradicted (Nielsen, 2014) by the data used during the development of this theory but never properly analysed does not contribute to the advancement of science but only to the advancement of dubious ideas and to the irrelevant interpretations of the mechanism of growth.

In contrast, the process of *deriving* mathematical formulae starts with the well-defined assumptions and follows closely the mathematical chain of logical reasoning. The final formulae might be complicated but they are still acceptable if they are based on the simplest possible starting assumptions. We can question the original assumptions but we have no problem with understanding why the final formula is expressed in a certain specific way because each step in deriving such a formula has been mathematically justified. We do not have such an assurance with constructed formulae because they come from nowhere.

Constructing complicated mathematical expressions is not recommended in scientific research but looking for the *simplest* mathematical representation of data is justified because it is far better to base the interpretation of data on the mathematical analysis than on pure imagination or at best on a superficial examination of data (e.g. Ashraf, 2009; Galor, 2005a, 2005b, 2007, 2008a, 2008b, 2008c, 2010, 2011, 2012a, 2012b, 2012c; Galor and Moav, 2002; Snowdon & Galor, 2008).

When looking for a mathematical representation of data we do not have to derive mathematical formulae starting from certain simple assumptions. We can use simple but well known mathematical distributions, such as a straight line or an exponential function. For instance, hyperbolic interpretation of growth is based on the linear interpretation of the reciprocal values of data. After finding the simplest description of data the next step is then to try to explain why the data follow a certain, mathematically-defined distribution. We are then dealing with finding the mechanism of growth.

The analysis based on the examination of *growth rates* also starts with finding their simplest mathematical descriptions but they have to be then translated into the mathematical descriptions of the *size* of the growing entity. Results of such translations might be complicated but they are still acceptable if they are based on the simplest mathematical descriptions of the growth rate because the ultimate aim of explaining the mechanism of growth is then also simplified. We shall not have to explain why the size of the growing entity is described by a complicated mathematical distribution but rather why the growth rate can be described in a certain simplest way. For instance, we shall see that the linearly-decreasing growth rate generates a *non*-linear growth, which is described by a fairly-complicated mathematical formula. The aim of explaining the associated mechanism is then reduced to explaining why the growth rate decreases linearly.

If carried out properly, mathematical analysis of data is an important and essential step in scientific investigations. The commonly repeated mistake that population was or is increasing exponentially is based on neglecting the analysis of data. When, properly analysed, data demonstrate that the population was never increasing exponentially. The same applies to the *historical* economic growth (Nielsen, 2015a). *Modern* economic growth can be *approximated* by using exponential function over a certain range of time (Nielsen, 2014) but other trajectories might be also possible. The popular habit of describing any type of growth as exponential is unjustified because there are many other types of growth.

Close analysis of data is essential in explaining the mechanism of growth, because the data can help to eliminate various irrelevant mechanisms. For instance, the past economic growth and the growth of human population can be described using the simple first-order hyperbolic distributions (Nielsen, 2014; von Foerster, Mora, & Amiot, 1960). Such information is already a significant step forward because rather than looking for a variety of possible mechanisms of growth we can now focus on trying to explain why the historical growth was hyperbolic.

We might also take an alternative approach based on the examination of the empirical growth rates. The growth rate for the hyperbolic growth is directly proportional to the size of the growing entity. So now we do not even have to ask why the growth was hyperbolic, the question, which might be difficult to answer because hyperbolic distributions appear to be complicated, but to ask why the growth rate is directly proportional to the size of the growing entity, which might be easier to answer because linear trends are much easier to understand than the more complicated hyperbolic trends.

The direct analysis of $S$ or $F(S)$ is simple and does not require any further explanation. We shall now concentrate on the analysis of the growth rate, $R$, of the growing entity $S$ or the growth rates of the appropriately-defined functions $F(S)$, such as $F(S) = \ln S$. The general concept is to reproduce the growth rates using the simplest mathematical distributions. We shall show how such simple mathematical representations of growth rates can be converted to mathematical expressions describing the growth of a studied process. The derived mathematical formulae might not appear simple but they will be based on the simplest starting steps. Mathematical representation of data based on such procedures can then be used not only to describe the existing data but also to predict future growth.

No prediction is absolutely reliable even if we know the mechanism of growth because the mechanism of growth can change. We can only predict the future on the assumption that the mechanism will remain unchanged and that the past pattern will be reflected in the future growth, but even then we might have more than one future pathway. Nevertheless, such predictions, based on rigorous analysis of data, could be useful. For instance, if we can demonstrate a strong probability of an exponential economic growth we could take such a prediction as a warning sign because exponential growth increases indefinitely and at a certain time it becomes unsustainable. Other types of growth might also become unsustainable and it is useful to explore such possibilities.

**Mathematical methods**

Growth rate is defined by the following equation:

$$R \equiv \frac{1}{S}\frac{dS}{dt} \approx \frac{1}{S}\frac{\Delta S}{\Delta t} \qquad (1)$$

where $S(t)$ is the size of the growing entity and $t$ is the time.

More explicitly, for the direct calculations from data:

$$R_{i+1} = \frac{1}{S_i}\frac{S_{i+1}-S_i}{t_{i+1}-t_i}. \qquad (2)$$

The size $S$ can represent, for instance, the Gross Domestic Product (GDP) or the size of the population. However, the described mathematical methods have a general application. In principle, they can be used for any type of growth. If we have a sufficiently large number of data, we can use them to determine the empirical growth rate, analyse it mathematically, use its mathematical description to fit the data and to predict growth.

If analysis of data is carried out by using an appropriately-defined distribution $F(S)$, rather than $S$, then the starting point is to calculate of the growth rate of $F(S)$:

$$R \equiv \frac{1}{F(S)}\frac{dF(S)}{dt} \approx \frac{1}{F(S)}\frac{\Delta F(S)}{\Delta t}. \qquad (3)$$

The simplest way to calculate the growth rate of $S$ is directly from data using the eqn (2). However, such calculations are sensitive to local gradients $\Delta S/\Delta t$ and consequently, they usually produce strongly fluctuating growth rates, which obscure the trend. A far better way is to calculate the growth rate by using the polynomially-interpolated gradient $\Delta S/\Delta t$. This method allows for a cleaner representation of the growth rate.

It should be noted that in fitting data and in predicting growth, strong fluctuations in the growth rate have no impact on the size of the growing entity (Nielsen, 2015b). Likewise, even sizable variations or oscillations in the growth rate usually have negligible effect. The trajectory of the growing entity is not determined by such variations but by *the general trend* of the corresponding growth rate. As an example, we are showing the data for Sweden and the results of their analysis using a constant rate of natural increase (RNI).

The top section of Figure 1 shows birth and death rates as well as the rate of natural increase, which in this case is approximately the same as the growth rate because the migration rates were small. The lower part of this figure shows the corresponding growth of the population. It is clear that even large fluctuations in the RNI or in the growth rate are not reflected in the growth of the population. These fluctuations can be neglected.

Furthermore, as shown in the top section of Figure 1, the RNI (or the growth rate) does not fluctuate around a constant value. Nevertheless, a constant value, which generates exponential growth, reproduces the data reasonably well. The clear disagreement between the calculated exponential growth and the data is towards the end of the displayed distribution when the RNI (or the growth rate) departs far from the assumed constant value. More data would have to be included to study this feature. It could be just a temporary deviation but it also could be a diversion to a new trend.

Our aim was not to analyse the data for Sweden but to demonstrate that even simple approximations of the growth rate trajectories could be successful in describing the growth of the population or the growth of any other growing entity. Consequently, we can neglect not only the fluctuations of the growth rate but also its periodic, longer-term, variations and we can use only the general trend of the growth rate to study growth trajectories and to calculate future trends.

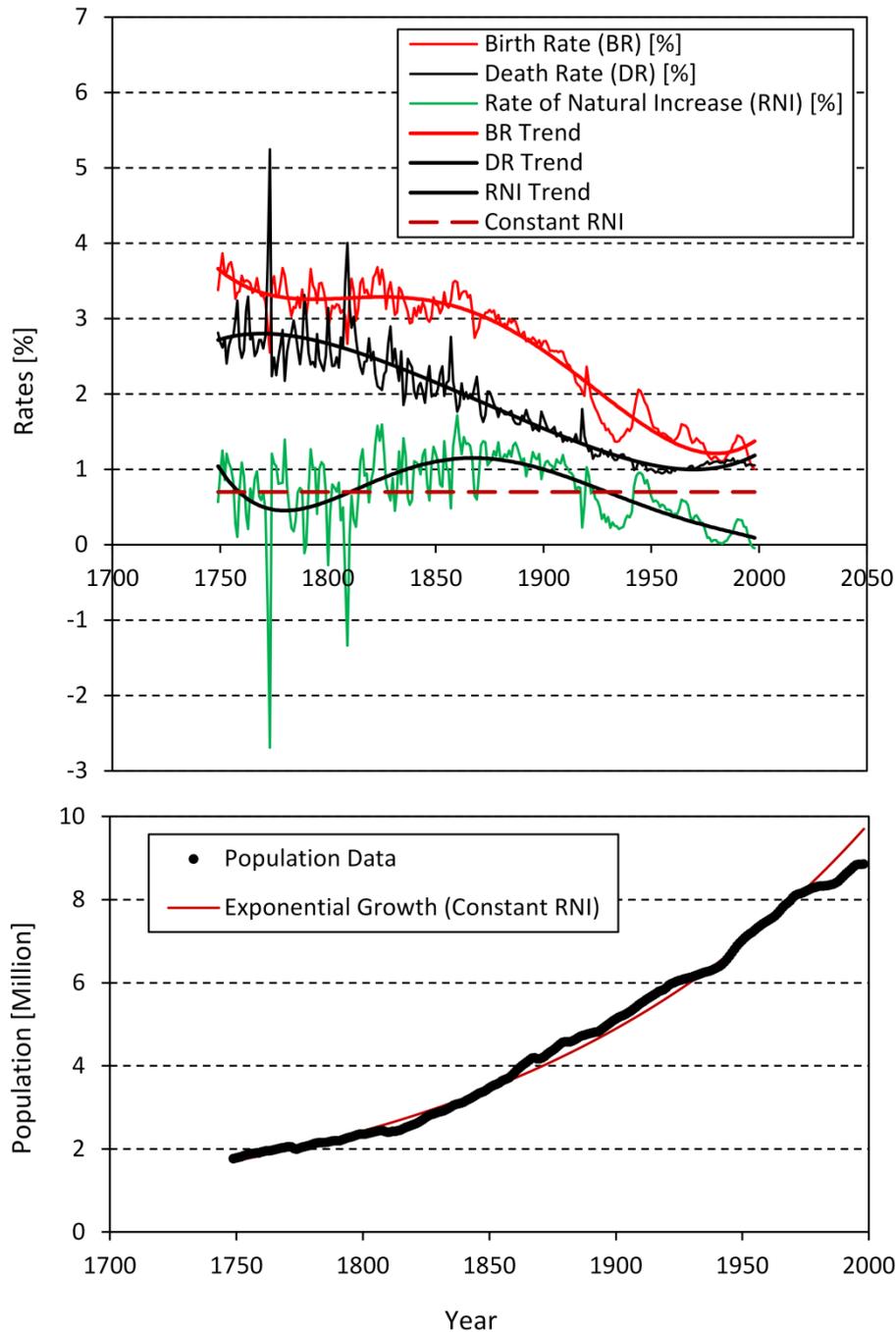

Figure 1. Demographic information for Sweden (Statistics Sweden, 1999).

Growth rate can be presented as a function of time or as a function of the size of the growing entity. We shall now discuss these two possibilities.

If the empirically-defined growth rate can be described by a certain time-dependent function $f(t)$, i.e. if

$$\frac{1}{S}\frac{dS}{dt} = f(t), \qquad (4)$$

then to find the mathematical representation of data we have to solve the following differential equation:

$$\frac{dS}{S} = f(t)dt. \qquad (5)$$

Its solution is

$$S(t) = \exp\left[\int f(t)dt\right]. \qquad (6)$$

If

$$f(t) = r = const, \qquad (7)$$

the solution of the eqn (4) is given by the exponential function,

$$S(t) = Ce^{rt}, \qquad (8)$$

where $C$ is related to the constant of the integration. The eqn (8) describes exponential increase, if $r > 0$ or decrease if $r < 0$.

If the empirically-determined growth rate can be represented by a straight line, i.e. if

$$f(t) = a + bt, \qquad (9)$$

where $a$ and $b$ are constants, then

$$S(t) = C\exp\left[at + 0.5bt^2\right]. \qquad (10)$$

In this case, the gradient of $S(t)$ is

$$\frac{dS(t)}{dt} = C(a + bt)\exp\left[at + 0.5bt^2\right]. \qquad (11)$$

If both $a$ and $b$ are positive, the size $S(t)$ will continue to increase indefinitely. However, if the fitted straight line is decreasing, i.e. if $b < 0$ while $a > 0$, then $S(t)$ will reach a maximum at $t = -a/b$ and it will then start to decrease.

Let us now assume that the empirically-determined growth rate can be expressed as a function of the size of the growing entity, $S$:

$$\frac{1}{S}\frac{dS}{dt} = f(S). \qquad (12)$$

We can express this equation as

$$\frac{dS}{S \cdot f(S)} = dt. \qquad (13)$$

We now have a mathematically more complicated problem, because there is no single prescription for the solution of such differential equations.

In the simplest case when $f(S) = r = const$ the solution is again represented by an exponential function. If we take the next least complicated step and assume that $f(S)$ is represented by a straight line, i.e. if

$$f(S) = a + bS, \tag{14}$$

then we have the following differential equation:

$$\frac{dS}{S\ a+bS} = dt. \tag{15}$$

To find how to integrate the left-hand side of this equation let us consider a general case:

$$\frac{dx}{(a+bx)(c+ex)}, \tag{16}$$

where $a$, $b$, $c$ and $e$ are constants.

To integrate this fraction we split it into two fractions:

$$\frac{1}{(a+bx)(c+ex)} = \frac{A}{(a+bx)} + \frac{B}{(c+ex)}, \tag{17}$$

where $A$ and $B$ are certain constants, which we now have to determine.

The right-hand side of the eqn (17) can be expressed as

$$\frac{A}{(a+bx)} + \frac{B}{(c+ex)} = \frac{(c+ex)A + (a+bx)B}{(a+bx)(c+ex)}. \tag{18}$$

By comparing the eqns (17) and (18) we can see that

$$(c+ex)A + (a+bx)B = 1, \tag{19}$$

which gives us a set of two equations:

$$cA + aB = 1, \tag{20a}$$

$$eA + bB = 0. \tag{20b}$$

Their solution is

$$A = \frac{b}{\Delta}, \tag{21a}$$

$$B = -\frac{e}{\Delta}, \tag{21b}$$

where

$$\Delta = cb - ae. \tag{22}$$

So now, the eqn (17) can be replaced by

$$\frac{1}{(a+bx)(c+ex)} = \frac{b}{\Delta}\frac{1}{(a+bx)} - \frac{e}{\Delta}\frac{1}{(c+ex)}. \tag{23}$$

The integration of the left-hand side of this equation is replaced by the integration of two simpler fractions. Their integration can be done by substitutions. Thus, for instance if we use $u = a + bx$ we get

$$\int \frac{1}{a+bx} dx = \frac{1}{b} \int \frac{du}{u} = \frac{1}{b} \ln u = \frac{1}{b} \ln(a+bx). \quad (24)$$

Consequently,

$$\int \frac{dx}{(a+bx)(c+ex)} = \frac{1}{\Delta} \ln \frac{a+bx}{c+ex}. \quad (25)$$

We have derived a useful general formula of integration. In particular, we can see now that

$$\int \frac{dx}{x(a+bx)} = -\frac{1}{a} \ln \frac{a+bx}{x}, \quad (26)$$

because $c = 0$, $e = 1$ and consequently $\Delta = -a$.

We are now ready to solve the eqn (15). The integration of both sides of the equation

$$\int \frac{dS}{S \; a+bS} = \int dt \quad (27)$$

gives

$$-\frac{1}{a} \ln \frac{a+bS}{S} = t + C, \quad (28)$$

where $C$ is the constant of integration.

Simple arithmetical manipulations lead to the following solution of the eqn (15):

$$S = \left[ Ce^{-at} - \frac{b}{a} \right]^{-1}. \quad (29)$$

The constant $C$ can be determined by normalising calculated $S$ to data at a certain time $t_0$,

$$C = \frac{a+bS_0}{S_0 a e^{-at_0}}, \quad (30)$$

where $S_0$ is the empirical size of the growing entity (e.g. the GDP) at a selected time $t_0$.

If $a + bS = r = const$, i.e. if the growth rate is constant, the eqn (29) gives exponential growth.

If $a + bS \neq const$ we have two possibilities: the growth rate represented by $a + bS$ can either increase or decrease with the size of the growing entity:

$$\frac{1}{S} \frac{dS}{dt} = a + bS. \quad (31)$$

If $b < 0$, the eqn (29) represents the logistic-type of growth. The characteristic signature of this type of growth is its linearly decreasing growth rate. The corresponding size $S$ of the growing entity approaches asymptotically a maximum value of

$$S_\infty = \frac{a}{|b|}. \qquad (32)$$

The eqn (32) defines the mathematical *limit to growth*, which is often described as the *carrying capacity* but it is only the carrying *capacity* if parameters *a* and *b* are clearly and convincingly related to the well-defined and well-explored ecological limits; otherwise, the calculated limit $S_\infty$ is just the calculated limit to growth, which may or may not represent the carrying capacity.

For instance, if we consider the growth of the GDP and if we determine empirically the parameters *a* and *b* using the *empirical values* of the growth rate it would be incorrect to claim that the calculated $S_\infty$ represents the empirically determined *carrying capacity* because the past economic growth might be following an unsafe trajectory and the economic collapse might happen even before reaching the calculated limit $S_\infty$. For this reason, describing the logistic limit as the carrying capacity may be misleading and it would be perhaps better to avoid such descriptions.

The same comment applies also the calculated maximum when using the eqn (10). The calculated maximum, even if based on using the empirically-determined parameters *a* and *b*, is just the calculated maximum. It also does not describe the carrying *capacity*. With limited resources the growth might be terminated even before reaching the maximum calculated using the empirically-determined parameters.

If $b > 0$ then, according to the eqn (29), the growth approaches singularity (escapes to infinity) at the time

$$t = t_s = -\frac{1}{a}\ln\frac{b}{aC}. \qquad (33)$$

This type of growth resembles the hyperbolic growth, which characterises the historical economic growth and the growth of human population (Nielsen, 2014, 2015a; von Foerster, Mora, & Amiot, 1960). Hyperbolic growth (or to be more precise, the first-order hyperbolic growth) is given by the following simple equation:

$$S = (C - bt)^{-1}, \qquad (34)$$

where $b > 0$.

Hyperbolic growth escapes to infinity when

$$t = t_s = \frac{C}{b}. \qquad (35)$$

Hyperbolic distribution is a solution of the following differential equation:

$$\frac{1}{S}\frac{dS}{dt} = bS. \qquad (36)$$

If we compare this equation with the eqn (31) we can see that they are similar. In both cases, growth rate changes linearly with the size of the growing entity. However, while for the hyperbolic growth [eqn (36) with $b > 0$] the growth rate is *directly* proportional to *S*, for the growth described by the eqn (31) the linearly-changing growth rate is displaced by the parameter *a*. It is a small difference but with significant consequences and it is important to understand the similarities and differences between these two different patterns of growth.

Their corresponding solutions given by eqns (29) and (34), for $b > 0$, are presented in Figure 2. Their reciprocal values, $1/S(t)$, are shown in Figure 3. The distribution described by the differential equation (31) and by its solution (29) for $b > 0$ is not hyperbolic but it escapes to infinity at a fixed time [see eqn (33)]. We could, therefore, call it a *pseudo-hyperbolic* distribution.

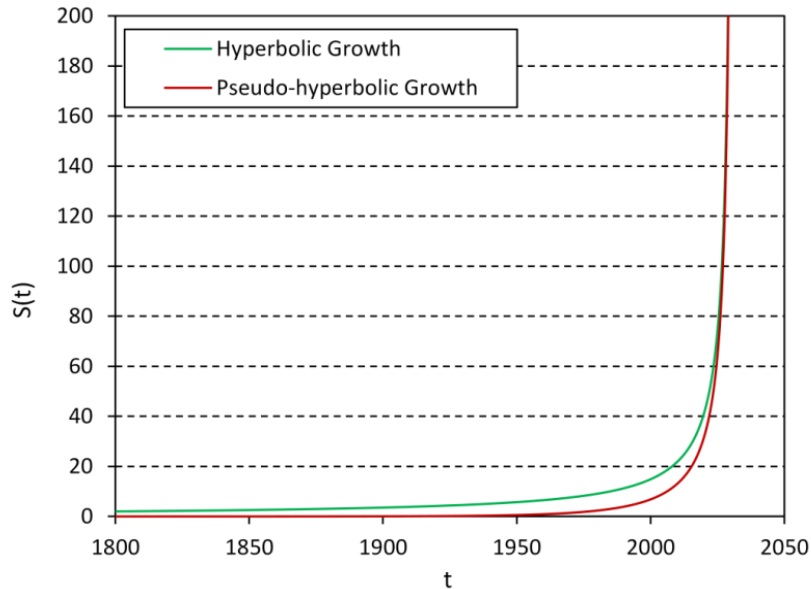

Figure 2. Comparing hyperbolic distribution given by the eqn (34) with the pseudo-hyperbolic distribution given by the eqn (29) for $b > 0$. In this example, they escape to infinity at the same time.

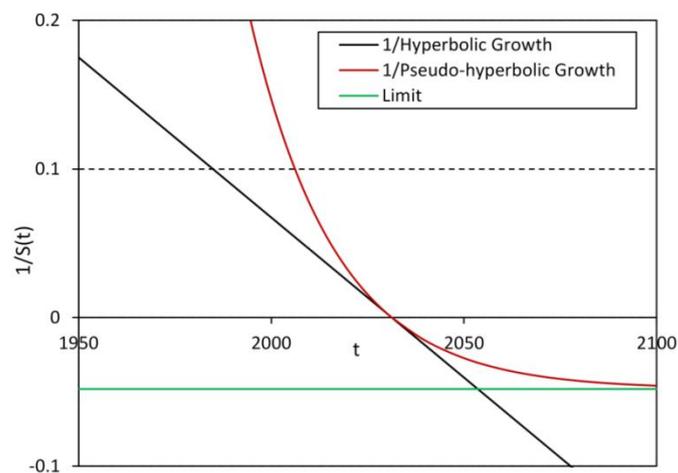

Figure 3. Comparing the reciprocal values, $1/S(t)$, of the hyperbolic distribution given by the eqn (34) of the pseudo-hyperbolic distribution given by the eqn (29) for $b > 0$. In this example, they cross the horizontal axis at the same time. The limit for the reciprocal values of the pseudo-hyperbolic distribution is $-b/a$.

The curious difference between the respective differential equations, (31) and (36), is that the eqn (31) cannot be treated as the generalisation of the eqn (36). The two equations have to be solved independently. The solution to the eqn (31) cannot be used to derive the solution to the eqn (36). While solving the eqn (31) is difficult, solving the eqn (36) is simple. The solution can be obtained by substitution $S = Z^{-1}$.

Both distributions, the hyperbolic distribution described by the eqn (34) and the pseudo-hyperbolic distribution described by the eqn (29), escape to infinity at a certain time but the essential difference is in the behaviour of their reciprocal values, $1/S$ (see Figure 3). For the hyperbolic growth, the reciprocal values decrease linearly with time. For the pseudo-hyperbolic growth, given by the eqn (29), they decrease non-linearly approaching the limit of $-b/a$.

In the example shown in Figures 2 and 3, parameters for the pseudo-hyperbolic growth are: $a = 4.475 \times 10^{-2}, b = 2.155 \times 10^{-3}$ and $C = 1.437 \times 10^{38}$. Parameters for the hyperbolic growth are $b = 2.155 \times 10^{-3}$ and $C = 4.376 \times 10^{0}$. Singularity is at $t = 2031.35$.

A summary of the discussed differential equations, their solutions and their properties is presented in Table 1.

Fitting data and projecting growth can be also carried out by replacing the growth rate of $S$ by the growth rate of any, suitably-defined, distribution $F(S)$. The aim here is again to look for the simplest mathematical descriptions of growth rates. If the mathematical description of the growth rate of $S$ is complicated, it might be possible that the mathematical description of the growth rate of $F(S)$ could be simpler. Analysis of data can be simplified by looking for their alternative representations and the general idea is to try to reduce the analysis, if possible, to the simplest mathematical expression – the straight line.

**Table 1.** Summary of the discussed differential equations, their solutions and properties

| Differential Equation | Solution | Comments |
|---|---|---|
| $\dfrac{1}{S}\dfrac{dS}{dt} = r$ | $S = Ce^{rt}$ | Exponential increase (if $r > 0$) or decrease (if $r < 0$) |
| $\dfrac{1}{S}\dfrac{dS}{dt} = f(t)$ | $S = C\exp\left[\int f(t)dt\right]$ | |
| $\dfrac{1}{S}\dfrac{dS}{dt} = a + bt$ | $S = C\exp\left[at + 0.5bt^2\right]$ | If $b > 0$, growth increases indefinitely. If $b < 0$, growth reaches a maximum at $t = a/|b|$ |
| $\dfrac{1}{S}\dfrac{dS}{dt} = bS$ | $S = (C - bt)^{-1}$ | If $b > 0$, hyperbolic growth. Singularity at $t_s = C/b$. Reciprocal values, $1/S$, decrease linearly with time. |
| $\dfrac{1}{S}\dfrac{dS}{dt} = a + bS$ | $S = \left[Ce^{-at} - \dfrac{b}{a}\right]^{-1}$ | If $b > 0$: pseudo-hyperbolic growth. Singularity at $t_s = -\dfrac{1}{a}\ln\dfrac{b}{aC}$. Reciprocal values decrease non-linearly to $-b/a$ when $t \to t_\infty$. If $b < 0$: logistic growth. $S$ increases asymptotically to $a/|b|$ when $t \to t_\infty$. |

Thus, for instance, if $F \equiv \ln S$, where $S$ represents the empirically-determined size of the growing entity, and if

$$\frac{1}{F}\frac{dF}{dt} = a + bt, \qquad (37)$$

then

$$F = C\exp(at + 0.5bt^2), \qquad (38)$$

and

$$S = \exp\left[C\exp(at + 0.5bt^2)\right]. \qquad (39)$$

If $F \equiv \ln S$ and if

$$\frac{1}{F}\frac{dF}{dt} = a + bF, \qquad (40)$$

then

$$F = \left(Ce^{-at} - \frac{b}{a}\right)^{-1}, \qquad (41)$$

and

$$S = \exp\left[\left(Ce^{-at} - \frac{b}{a}\right)^{-1}\right]. \qquad (42)$$

Mathematical representations of $S$ given by the eqns (39) and (42) are not simple but they are acceptable because they are based on reducing mathematical analysis of data to the simplest representation given by a straight line for the growth rate of $F$.

We can also extend this alternative representations by replacing *the growth rate R* by a suitably defined function $F(R)$. If the mathematical description of the growth rate of $S$ turns out to be complicated it might be possible that a suitably-defined function $F(R)$ could simplify the analysis.

Thus for instance, visual examination of the empirical growth rate of $S$ might suggest that it depends hyperbolically on time. We might try to fit hyperbolic distribution to the empirically-determined growth rate but it is also a good idea to check whether the distribution is indeed hyperbolic by examining the reciprocal values of $R$ because if $1/R$ follows a straight line then $R$ is hyperbolic. If

$$F(R) = \frac{1}{R} = a + bt \qquad (43)$$

then

$$R = \frac{1}{S}\frac{dS}{dt} = \frac{1}{a + bt} \qquad (44)$$

Hyperbolic distribution is not as simple as a straight line but it can be reduced to a straight line, which is easier to accept and understand. Such an exercise increases confidence that the distribution is indeed hyperbolic or at least that it can be well approximated by a hyperbolic distribution.

The differential equation (44) can be presented as

$$\frac{dS}{S} = \frac{dt}{a+bt}, \tag{45}$$

which, when integrated, gives

$$\ln S = \frac{1}{b}\ln(a+bt) + C. \tag{46}$$

Consequently,

$$S = C(a+bt)^{1/b} \tag{47}$$

because $\exp(\ln z) = z$. The constants $C$ are different in these last two equations but it does not matter because they are just the normalisation constants, which have to be determined by comparing calculated $S$ with its corresponding empirical value.

The eqn (47) is not simple but it has been obtained by reducing mathematical analysis to the simplest mathematical expression given by the eqn (43), which identifies hyperbolic distribution of $R$. The fundamental starting step is simple and the derived expression for $S$, even if complicated, can be accepted with a high degree of confidence.

If a visual examination of the empirical growth rate $R$ suggests that it follows an exponential distribution we can try to fit an exponential function to $R$ or to display it using the semilogarithmic scales of reference. If

$$\ln R = a + bt, \tag{48}$$

then

$$\frac{1}{S}\frac{dS}{dt} = \exp(a+bt) \tag{49}$$

and the solution to this equation is

$$S = C\exp\left[\frac{e^a}{b}e^{bt}\right] \tag{50}$$

Again, it is not a simple description of $S$ but this complicated expression has been derived using the simplest representation of $R$ via $\ln R$.

All mathematical descriptions of $S$ presented here [see eqns (10), (29), (34), (39), (42), (47) and (50)] are not simple but all of them were derived using the simplest mathematical representations of related quantities. It is easy to *construct* complicated but dubious formulae but even complicated formulae are acceptable if they are *derived* using simple and acceptable assumptions.

**Example**

Application of the discussed methods is illustrated in Figures 3 and 4 using the world GDP data (World Bank, 2015). In Figure 4 we present two sets of calculations of the growth rate of the GDP: (1) the growth rate calculated directly from data and (2) the growth rate calculated using the interpolated gradient. As expected, the growth rate calculated directly from data is characterised by strong fluctuations. Such calculations do not help in unravelling the prevailing trend. However, the trend becomes clear if we calculate the growth rate using the interpolated gradient.

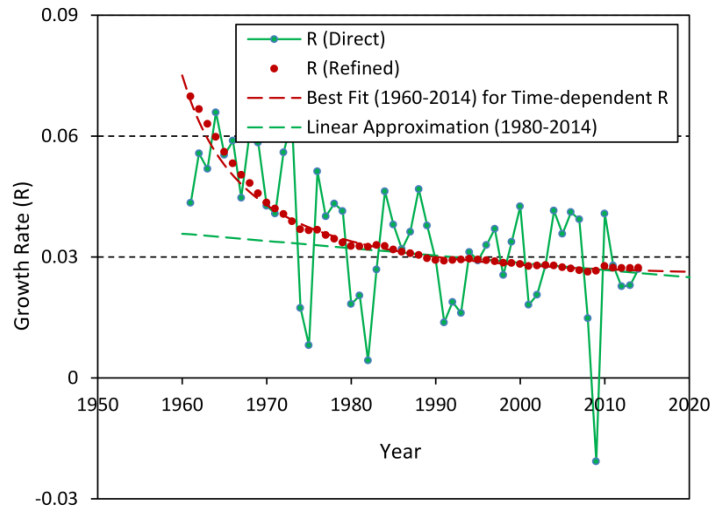

Figure 4. The growth rate, *R*, describing global economic growth between 1960 and 2014, calculated directly from data, *R* (*Direct*), and by using the interpolated gradient, *R* (*Refined*). Calculations were carried out using the GDP data of the World Bank (2015).

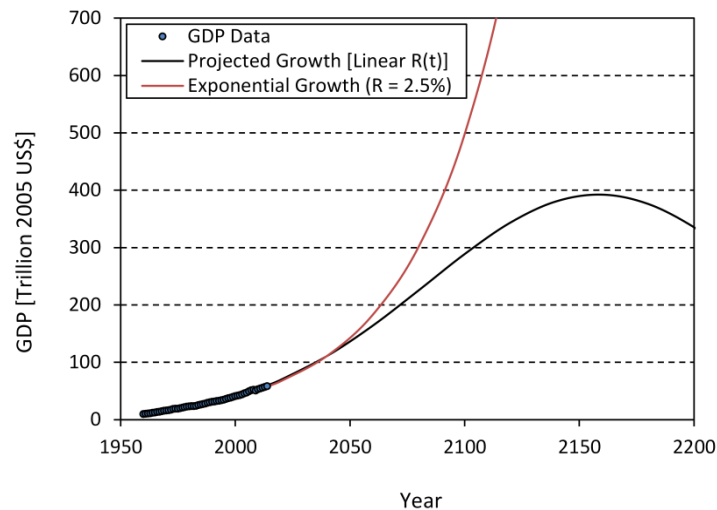

Figure 5. Fitted GDP data and the projected growth using linear approximation for the growth rate between 1980 and 2014 and a constant growth rate of 2.5% per year. Calculations were supported by the GDP data of the World Bank (2015).

The growth rate calculated using the interpolated gradient follows the reciprocal of a familiar mathematical distribution (Nielsen, 2011) labelled here as the best fit, which will be discussed in a separate publication. The growth rate of the world GDP approaches asymptotically a constant value of 2.5%, which is close to the value of 2.7% in 2014. The world economic growth increases already approximately exponentially.

Here we shall use simpler approximations: a gradually-decreasing straight line obtained by fitting empirical growth rate between 1980 and 2014, as shown in Figure 4, and a horizontal straight line representing exponential growth of the GDP.

The linear approximation of the time-dependent growth rate between 1980 and 2014 is described by the following parameters: $a = 3.895 \times 10^{-1}$ and $b = -1.805 \times 10^{-4}$. In this approximation, the growth rate decreases slowly to zero, so if we use these parameters in the eqn (10) we can expect that the future economic growth will reach a maximum and will start to decrease. However, the gradient of the straight line is small. Between 1980 and 2014 the growth rate decreased from 3.5% per year to only 2.7%. Exponential trajectory shown in Figure 5 was calculated using a constant asymptotic growth rate of 2.5%.

If the growth rate is going to remain constant at this value, the world GDP will increase to about $500 trillion of 2005 US$ by the end of the current century, providing that such an increase can be supported. However, if the growth rate is going to decrease linearly, as shown in Figure 4, then the growth of the world GDP is projected to reach a maximum of about $390 trillion around 2158, again providing that such large GDP can be ecologically supported.

**Summary and conclusions**

We have discussed mathematical methods of growth rate analysis and of predicting growth. In their simplest applications, the growth rate is presented either as a function of time or as a function of the size of the growing entity. The aim is then to find the simplest mathematical description of the growth rate and to use it to calculate growth trajectories. In the extended application we can define a quantity $F(S)$ which depends on the size of the growing entity and we can apply the same methods by examining the time dependence or the size dependence of $F$.

The described methods are an extension to the usual analysis of the size $S$ of the growing entity or of the analysis of the appropriately defined distributions depending on $S$, such as $F(S) = 1/S$ or $F(S) = \ln(S)$. These methods are illustrated using the world GDP data but they can have more general applications.